\documentclass[twocolumn,aps,floatfix,showpacs,showkeys]{revtex4}
\usepackage{color,graphicx}
\usepackage{dcolumn}
\usepackage{bm}
\usepackage{epsfig}
\usepackage{supertabular}
\usepackage[bookmarksopen]{hyperref}
\def\be {\begin{equation}}
\def\ee {\end{equation}}
\def\nn {\nonumber}
\def\bea {\begin{eqnarray}}
\def\eea {\end{eqnarray}}

\begin{document}
\title{Collective modes of an anisotropic quark-gluon plasma induced by relativistic jets}
\bigskip
\bigskip
\author{ Mahatsab Mandal}
\email{mahatsab.mandal@saha.ac.in}
\author{ Pradip Roy}
\email{pradipk.roy@saha.ac.in}
\affiliation{Saha Institute of Nuclear Physics, 1/AF Bidhannagar
Kolkata - 700064, India}

\begin{abstract}
We discuss the characteristics of collective modes induced by relativistic 
jets in an anisotropic quark-gluon plasma(AQGP).
Assuming 
a tsunami-like initial jet distribution, 
it is found that the dispersion relations for both the stable and 
unstable modes are modified substantially 
due to the passage of jet compared to the case when there is no jet. It has also
been shown that the growth rate of instability first increases compared 
to the no jet case and then completely turned into damping 
except the case when the jet velocity is perpendicular to the wave vector in 
which case the instability always grows. Thus, the introduction
of the jet in the AQGP, in general, might to faster isotropization 
for the special case when the wave vector is parallel to the anisotropy axis.
\end{abstract}

\maketitle

\section{Introduction}
The primary goal of the ultra-relativistic heavy-ion collision experiments at BNL RHIC and at CERN LHC is 
to study the properties of a deconfined state of the QCD matter, commonly known as quark-gluon plasma(QGP). 
According to lattice calculation, the novel state of matter is expected to be formed when the temperature of the nuclear 
matter is raised to the critical value $T_c\sim170$MeV, or the energy density of the nuclear matter is raised to above 1 GeV/$fm^3$.
High energy partons behave as hard probes which are produced in the early stage of the collision due to hard scattering. 
In relativistic heavy-ion collision jets with high transverse momentum travels through the hot and dense medium and it losses 
energy by collisional(interaction with thermal quark and gluon) and  radiative processes(bremsstrahlung).
This phenomena  is commonly known as jet quenching, because in the direction of propagation of the jet one observes
a decrease of high energy hadrons and increase in the number of soft hadrons. In addition, the passage of jet influences
the collective modes of the system as we shall see below.


In Refs~\cite{prd74,prd76,prd77}, the effect of a jet of particle on the plasma properties, propagated through 
the equilibrated and isotropic QGP has been analyzed, where initial configuration of jet is 
assumed to be colorless, electrically neutral and described by tsunami-like distribution~\cite{hep-ph971} 
in momentum space. Non-equilibrium jet of particles while traveling through QGP  destabilizes the plasma 
producing the collective gauge fields. These gauge modes might have several branches and some of these might even be unstable.
The most important among those are the modes which grow exponentially in time. When the instability occurs, the kinetic 
energy of the  particles is converted to the field energy  which  speeds up the equilibration process
and also leads to  faster isotropization~\cite{prc49,prl94,prl94A,appb} of the QGP. The main assumption for 
such phenomenon to occur is that the time to generate the growth of the gauge fields is smaller than the hadronization time.
Plasma instabilities fully develop on the time scale of the order of $t\sim(6.7-12.5)/\omega_t$\cite{prd76,prd77}, where 
$\omega_t$ is the total angular frequency of the whole system. 
In a weak coupling regime, the time scale  is $1-2$ fm/c  at $T\sim350$MeV.

In the present work we shall concentrate on the collective modes in AQGP induced by relativistic jets.
In the early stage of heavy-ion collision due to rapid longitudinal expansion, 
the plasma cools faster in the longitudinal direction 
leading to $\langle p_L^2\rangle<<\langle p_T\rangle$. Such momentum-space 
anisotropy  leads to  collective modes having characteristic behavior distinct from what happens in isotropic plasma 
which has been extensively studied in ~\cite{prd68, prd70} where it is shown that the gluonic collective modes can be unstable. 
When a relativistic stream of particles interact with the non-equilibrium  plasma, with an anisotropic distribution
in momentum space, the behavior of collective modes  change as will be demonstrated in the following. 

In studying the evolution  of such a system, we use the method of the plasma physics within the frame work of the quark-gluon
transport theory~\cite{prc49,pr183} in weak coupling regime, i.e. $g << 1$. The time scale for the evolution of collective
modes is assumed to be much shorter than the inter-particle collision time. In this approach we have neglected 
hard mode interactions, assuming that the interactions between jet and plasma is only mediated by mean gauge fields. 

Kinetic instability can occur due to the interaction of plasma and jet partons. They are initiated either by charge 
or current fluctuation. In  the first case, the electric field is longitudinal i.e. the field is parallel to the wave vector 
${\bf k}$ (${\bf E}||{\bf k}$), while in the second case the field is perpendicular to ${\bf k}$ (${\bf E} \perp {\bf k}$). 
For this reason the corresponding instabilities are called longitudinal and 
transverse instabilities respectively. 
Since the electric field plays a crucial role in the generation of longitudinal modes, they are also called electric modes, 
while the transverse modes are called magnetic modes. The magnetic mode known as filamentation- or Weibel-instability~\cite{prl2} 
appears to be relevant for the QGP~\cite{prc49,appb}. In momentum-space anisotropic plasma, the growth rate of magnetic 
instability is maximum in the direction of anisotropy~\cite{prd68,prd70}.  
So we concentrate on this special case in which the momentum of the collective mode is in the direction of the anisotropy.
We expect that the dispersion relations will be affected due to the passage of jet in AQGP. In fact, we shall see 
that the growth rate of unstable modes first increases compared to the case of no jet scenario and then becomes damped.

The organization of the paper is as follows. In section 2 we briefly recapitulate Vlasov-type transport equation for
the colored particles. In section 3 and 4 we shall  show how the dispersion relations are modified in the presence of 
a tsunami like jet. Section 5 will be devoted to discuss the results followed by summary in section 6.


\section{Vlasov-type Transport equation for the colored particles}
Transport theory provides a natural framework to study equilibrium and non equilibrium plasmas. 
We review the transport equations obeyed by the distribution of colored particles and the classical
fields.
The (anti)quark distribution functions $f(p,x)$ and $\bar{f}(p,x)$, which are $N_c\times N_c$ hermitian matrices,
belong to the fundamental representation of the $SU(N_c)$ group. The gluons are the adjoint representation of $SU(N_c)$, 
and their distribution function ${\cal G}(p,x)$ is a $(N_c^2-1)\times(N_c^2-1)$ matrix in color space. The partials of jet  
consisting only quarks, and the corresponding distribution function can be described by a $N_c\times N_c$ matrix in
color space as $W_{jet}(p,x)$. 

In the fundamental representation, the color four-current $j^{\mu}$ is expresses as 
\bea
j^{\mu}(x)=&-&\frac{g}{2}\int_p p^{\mu}\Big[f(p,x)-{\bar f}(p,x)\nonumber\\
&-&\frac{1}{N_c}{\rm Tr}[f(p,x)-{\bar f}(p,x)]\nonumber\\
&+&2\tau^a{\rm Tr}[T^a{\cal G}(p,x)]\Big]\label{current}
\eea
where $\tau^a,T^a$ with $a = 1,...N_c^2-1$ are the $SU(N_c)$ group generator in the fundamental and adjoint 
representations with ${\rm Tr}[\tau^a\tau^b]=\frac{1}{2}\delta^{ab}$, ${\rm Tr}[T^aT^b]=N_c\delta^{ab}$ and 
$g$ is the QCD coupling constant. Here we use the notation
\be
\int_p...\equiv \int \frac{d^4p}{(2\pi)^3}2\Theta(p_0)\delta(p^2)
\ee

The transport equations of the quarks, antiquarks, gluons and the jet particle read as~\cite{pr183}
\bea
p^{\mu} D_{\mu}f(p,x) + \frac{g}{2}p^{\mu}\{F_{\mu\nu}(x),\partial^{\nu}_pf(p,x)\} = C,\label{t1}\\
p^{\mu} D_{\mu}\bar{f}(p,x) - \frac{g}{2}p^{\mu}\{F_{\mu\nu}(x),\partial^{\nu}_p\bar{f}(p,x)\} = \bar{C},\label{t2}\\ 
p^{\mu}{\cal D}_{\mu}{\cal G}(p,x) + \frac{g}{2}p^{\mu}\{{\cal F}_{\mu\nu}(x),\partial^{\nu}_p{\cal G}(p,x)\} = C_g,\label{t3}\\
p^{\mu}D_{\mu}W_{jet}(p,x) + \frac{g}{2}p^{\mu}\{F_{\mu\nu}(x),\partial^{\nu}_pW_{jet}(p,x) = C_W\label{t4}
\eea
where ${\cal F}_{\mu\nu}=\partial_{\mu}{\cal A}_{\nu}-\partial_{\nu}{\cal A}_{\mu}-ig[{\cal A}_{\mu},{\cal A}_{\nu}]$ is
the gluon field strength tensor, while $F_{\mu\nu}$ represents the strength tensor in the fundamental representation and  
the covariant derivatives $D_{\mu}$ and ${\cal D}_{\mu}$ act as\\
$~~~~~~~~~~~~~D_{\mu}=\partial_{\mu}-ig[A_{\mu}(x),...]$,\\
$~~~~~~~~~~~~~{\cal D}_{\mu}=\partial_{\mu}-ig[{\cal A}_{\mu}(x),...]$,\\
with gauge field $A_{\mu}=A^{\mu}_a(x)\tau^a$ and ${\cal A}_{\mu}=A^{\mu}_aT^a$.

Now, we will  study the Vlasov approximation, or collisionless dynamics of the color conductor where 
the mean free  time of the collisional processes  are lager then the  time scale of evolution of the whole system. 
Therefore the collision terms $C, {\bar C}, C_g$ and $C_W$ are neglected. The characteristic time scale~\cite{prd59} of the system's
evolution due to interparton collision is $t_{hard}\sim(g^4T\ln(1/g))^{-1}$ and  $t_{soft}\sim(g^2T\ln(1/g))^{-1}$.
So our current approximations are really only valid for time intervals shorter then $t_{soft}$.

In this work we will also assume that initially the plasma is colorless and in thermally equilibrated state.
Considering the distribution functions are deviation 
from their equilibrium states we have,
\bea
f(p,x)=n_{fd}({\bf p})+\delta f(p,x),\nonumber\\
{\bar f}(p,x)={\bar n}_{fd}({\bf p})+\delta {\bar f}(p,x),\nonumber\\
{\cal G}(p,x)=n_{be}({\bf p})+\delta {\cal G}(p,x),
\eea
where
\be
n_{fd/be}({\bf p})=\frac{1}{e^{{\bf p}/T}\pm1}
\ee
are the Fermi-Dirac and Bose-Einstein equilibrium distribution functions.

We also consider color fluctuation of the initial colorless distribution function of the jet. Thus
\be
W_{jet}=f_{jet}({\bf p})+\delta W_{jet}(p,x) 
\ee
where $f_{jet}(p)$ is the initial jet distribution function which we will consider to be 
of colorless tsunami-like form~\cite{hep-ph971}
\be
f_{jet}({\bf p})={\bar n}{\bar u}^0\delta^{(3)}({\bf p}-\Lambda {\bar {\bf u}}).\label{td}
\ee
Here ${\bar n}$ is a parameter proportional to the density; and ${\bar u}^{\mu}$ is the four-velocity. The parameter 
$\Lambda$ fixes the scale of energy of particles.


With the linear approximation of transport equations(Vlasov approximation) 
one can solve the polarization tensor for particles species $\alpha$~\cite{prd76,prd77,prd68,prd73}:
\be
\Pi^{\mu\nu}_{\alpha}(k)=g^2\int_p p^{\mu}\frac{\partial f_{\alpha}(\bf p)}{\partial p_{\beta}}
\Big(g^{\beta\nu}-\frac{p^{\nu}k^{\beta}}{p.k+i\epsilon}\Big)\label{dt}
\ee
where $\alpha$ specify the quarks, antiquarks, gluons or partials of jet. This tensor is symmetric, $\Pi^{\mu\nu}(k)=\Pi^{\nu\mu}(k)$,
and transverse, $k^{\mu}\Pi^{\mu\nu}=0$.

\section{ Collective modes of anisotropic quark-gluon plasma and the jet}
To include the local anisotropy in the plasma, one has to calculate the gluon polarization tensor incorporating anisotropic 
distribution function of the particles. This subsequently can be used to construct HTL corrected gluon propagator which, in general, 
assumes very complicated form. Such an HTL propagator was first derived in~\cite{prd68} in the temporal-axial gauge.

The spacelike component of the self-energy tensor can be written as
\be
\Pi^{ij}_p(k)=-g^2\int \frac{d^3p}{(2\pi)^3}v^i\partial^lf({\bf p})\Big(\delta^{jl}+\frac{v^jk^l}{k.v+i\epsilon}\Big)
\ee
The phase-space distribution is assumed to be given by the following ansatz~\cite{prd68,prd70}:
\be
f({\bf p})=f_{\xi}({\bf p})=N(\xi)f_{iso}(\sqrt{{\bf p}^2+\xi({\bf p.\hat{n}})^2}).
\ee
Here $f_{iso}$ is an arbitrary isotropic distribution function. $N(\xi)$ is the normalization constant which is equal to $\sqrt{1+\xi}$,
${\bf \hat{n}}$ is the direction of anisotropy. The parameter $\xi$ is the degree of anisotropy parameter $(-1<\xi<\infty)$
and is given by $\xi=\frac{1}{2}\frac{\langle p^2_T\rangle}{\langle p_z^2\rangle}-1$.
Making a change of variable (${\tilde p^2}=p^2(1+\xi({\bf p.\hat{n}})^2)$) the spatial components can be written as
\be
\Pi^{ij}_p(k)=m_D^2\sqrt{1+\xi}\int \frac{d\Omega}{(4\pi)}\frac{v^l+\xi({\bf v.\hat{n}})n^l}{(1+\xi({\bf v.\hat{n}})^2)^2}
\Big(\delta^{jl}+\frac{v^jk^l}{(k.v+i\epsilon}\Big)
\ee
where
\be
m_D^2=-\frac{g^2}{2\pi^2}\int^{\infty}_0dpp^2\frac{df_{iso}(p^2)}{dp}
\ee
is the isotropic Debye mass which depends on $f_{iso}$.\\
The self energy, apart from four-momentum($k^{\mu}$), also depends on the anisotropic vector $(n^{\mu}=(1,{\bf n}))$
and $\Pi^{\mu\nu}_p$ can be cast in a suitable tensorial basis appropriate for anisotropic
plasma in a co-variant gauge in the following way~\cite{prd68,plb662}:
\be
\Pi^{ij}_p(k)=\alpha A^{ij}+\beta B^{ij}+\gamma C^{ij}+\delta D^{ij}
\ee
where 
\bea A^{ij}&=&\delta^{ij}-k^ik^j/{\bf k^2},\label{s1}\\
B^{ij}&=&k^ik^j/{\bf k^2},\label{s2}\\
C^{ij}&=&\tilde{n}^i\tilde{n}^j/\tilde{n}^2,\label{s3}\\
D^{ij}&=&k^i\tilde{n}^j+k^j\tilde{n}^i,\label{s4}
\eea
where $\tilde{n}^i=A^{ij}n^j$ which obeys ${\bf\tilde{n}.k}=0$.\\
Now $\alpha, \beta, \gamma$ and $\delta$ are determined by the following contractions:
\bea
k^i\Pi^{ij}k^j&=&{\bf k}^2\beta,\nonumber\\
\tilde{n}^i\Pi^{ij}k^j&=&\tilde{n}^2{\bf k}^2\delta,\nonumber\\
\tilde{n}^i\Pi^{ij}\tilde{n}^j&=&\tilde{n}^2(\alpha+\gamma),\nonumber\\
{\rm Tr}\Pi^{ij}&=&2\alpha+\beta+\gamma
\eea
The expressions for structure functions have been given in Ref.~\cite{prd68}.
In the isotropic limit, $\xi\rightarrow0,$ the structure function $\gamma$ and $\delta$ vanish and $\alpha$ and $\beta$ are
directly related to the transverse and longitudinal components of the polarization tensor of the plasma respectively.

The dispersion law for the collective modes of anisotropic plasma in temporal axial gauge  can be 
determined by finding the poles of propagator $\tilde{\Delta}^{ij}$
\be
\tilde{\Delta}^{ij}(k)=\frac{1}{[({\bf k}^2-\omega^2)\delta^{ij}-k^ik^j+\Pi_p^{ij}(k)]}\label{pole}
\ee
Substituting Eqs. (\ref{s1}-\ref{s4}) in  the  above equation and performing the inverse formula~\cite{prd68} one finds 
\bea
\tilde{\Delta}(k)=\tilde{\Delta}_A[{\bf A}-{\bf C}]
+\tilde{\Delta}_G[({\bf k}^2-\omega^2+\alpha+\gamma){\bf B}\nonumber\\
+(\beta-\omega^2){\bf C}-\delta {\bf D}] 
\eea
The dispersion relation for the gluonic modes in anisotropic plasma is given by the zeros of 
\bea
\tilde{\Delta}^{-1}_A(k)&=&k^2-\omega^2+\alpha=0,\label{da}\\
\tilde{\Delta}_G^{-1}(k)&=&(k^2-\omega^2+\alpha+\gamma)(\beta-\omega^2)-k^2\tilde{n}^2\delta^2=0.\label{dg}\nonumber\\
\eea
If we examine the propagators (\ref{da}) and (\ref{dg}) in the static limit($\omega\rightarrow0$), we find that there are 
three mass scales~\cite{prd68}: $m_{\alpha}$~ and $m_{\pm}$. In isotropic limit, 
$\xi\rightarrow0$, $m_{\alpha}^2=m^2_-=0$ and $m^2_+=m^2_{D}$.

The solutions of the above two equations depend on $m_D,~\omega,~{\bf k},~\xi$ and ${\bf \hat{k}.\hat{n}}=\cos\theta_n$.
For $\xi>0$ one finds that there are at most three stable and two unstable modes  which depend on
$\theta_n$ and for $\xi<0$, there are three stable modes but only one is unstable~\cite{prd68,prd70}. 

Now, we shall briefly recall the calculation of the polarization tensor induced by the tsunami-like 
momentum distribution of the jet given by Eq.(\ref{td}). Substituting Eq.(\ref{td}) in Eq.(\ref{dt}) one deduces
the following expression of the polarization tensor 
for the jet partons:
\bea
\Pi^{ij}_{jet}(k)=-\omega^2_{jet}\Big(\delta^{ij}+\frac{k^iv^j_{jet}+k^jv^i_{jet}}{\omega-{\bf k.v}_{jet}}-
\frac{(\omega^2-{\bf k}^2)v_{jet}^iv_{jet}^j}{(\omega-{\bf k.v}_{jet})^2}\Big),\nonumber\\
\eea
where $v_{jet}$ is the velocity of jet and $\omega^2_{jet}=\frac{g^2\bar{n}}{2\Lambda}$ is the plasma frequency of the jet.
The dispersion laws of the collective modes of the system due to jet are  determined by searching the poles of 
the propagator of Eq.(\ref{pole}) by replacing $\Pi^{ij}_p$ with $\Pi^{ij}_{jet}$ i.e. by finding the solution $\omega(k)$. 
\section{Collective modes of the composite system}
Now we  study  the collective modes of the system due to propagation of an energetic jet  in an anisotropic quark gluon plasma. 
In very short time regime where the Vlasov approximation is valid, the total polarization of the system is given by 
the sum of the  two polarization tensors: 
\be
\Pi^{\mu\nu}_t(k)=\Pi^{\mu\nu}_{p}(k)+\Pi^{\mu\nu}_{jet}(k)\label{pit}
\ee\\
The dispersion relation of the collective modes of  the total system can be determined by solving the equation
\be
{\rm det}[(k^2-\omega^2)\delta^{ij}-k^ik^j+\Pi^{ij}_{t}(k)]={\rm det}[\Delta^{-1}(k)]^{ij}=0.\label{dr}
\ee
 
The solution of the above equation depends on $|{\bf k}|,~|{\bf v}_{jet}|,~{\bf \hat{k}.\hat{v}}_{jet}=\cos\theta_{jet},~m_{D},~
\xi, {\bf \hat{k}.\hat{n}}=\cos\theta_n$ and also on
\be
\eta=\frac{\omega^2_{jet}}{\omega_t^2},
\ee
where
\be
\omega_t^2=\omega^2_{jet}+\frac{m_D^2}{3}
\ee
Now the polarization tensor due to of jet partons can be decomposed in the following way:
\be
\Pi^{ij}_{jet} = \alpha^{\prime}A^{ij} + \beta^{\prime}B^{ij} 
\ee
where
\bea
\alpha^{\prime}&=&-\frac{\omega_{jet}^2}{2(\omega-kv_{jet}\cos\theta_{jet})^2}\nn\\
&\times&\Big(2\omega^2+v^2_{jet}(k^2-\omega^2)+v_{jet}\cos\theta_{jet}\nn\\
&&(v_{jet}\cos\theta_{jet}(k^2+\omega^2)-4k\omega)\Big)\nn\\
\beta^{\prime}&=&\omega^2\omega^2_{jet}\frac{v^2_{jet}\cos^2\theta_{jet}-1}{\omega-k v_{jet}\cos\theta_{jet}}
\eea
Therefore $\Delta^{-1}(k)$ in terms of the tensorial  basis becomes
\be
\Delta^{-1}(k) = ({\bf k}^2-\omega^2+\alpha+\alpha^{\prime}){\bf A}+(\beta+\beta^{\prime}-\omega^2){\bf B}
+\gamma{\bf C}+\delta{\bf D}.
\ee
Thus, we obtain an expression for the effective propagator as,
\bea
\Delta(k)&=&\Delta_A{\bf A}+({\bf k}^2-\omega^2+\alpha+\alpha^{\prime}+\gamma)\Delta_G{\bf B}\nonumber\\
&+&[(\beta+\beta^{\prime}-\omega^2)\Delta_G-\Delta_A]{\bf C}-\delta\Delta_G{\bf D}
\eea
with
\bea
\Delta^{-1}_A &=& {\bf k}^2-\omega^2+\alpha+\alpha^{\prime},\label{c1}\\
\Delta^{-1}_G&=&({\bf k}^2-\omega^2+\alpha+\alpha^{\prime}+\gamma)(\beta+\beta^{\prime}-\omega^2)\nonumber\\
&-&{\bf k}^2\tilde{n^2}\delta^2.\label{c2}
\eea

In the following subsection, we will analyze the nature of the modes of the whole system.
\subsection{Stable modes}
Now the dispersion relation of the  collective modes of the composite system is given by the zeros of Eqs. 
(\ref{c1}) and (\ref{c2}).
The propagator have poles at real value $\omega>|{\bf k}|$. The dispersion relation for $A$-modes can be determined by 
finding the solution of the equation
\bea
\omega^2_A = {\bf k}^2+\alpha(\omega_A)+\alpha^{\prime}(\omega_A)~\label{as}
\eea
In case of $G$-modes we factorize $\Delta_G^{-1}$ as
\bea
\Delta_G^{-1}=(\omega^2-\omega^2_{G+})(\omega^2-\omega^2_{G-})
\eea
where 
\bea
\omega^2_{G\pm} = \frac{1}{2}(\bar{\omega}^2\pm\sqrt{\Omega^2+
4{\bf k}^2\tilde{n^2}\delta^2}),~\label{cm2}
\eea
and 
\bea
\bar{\omega}^2 = \alpha+\alpha^{\prime}+\beta+\beta^{\prime}+\gamma+{\bf k}^2,\nonumber\\
\Omega=\alpha+\alpha^{\prime}-\beta-\beta^{\prime}+\gamma+{\bf k}^2.~\label{cm3}
\eea
For real $\omega>|{\bf k}|$, the square root of Eq. (\ref{cm2}) is always positive.
Therefore, at most two  stable modes come from $G$ -modes.

\begin{figure}[t]
\begin{center}
\epsfig{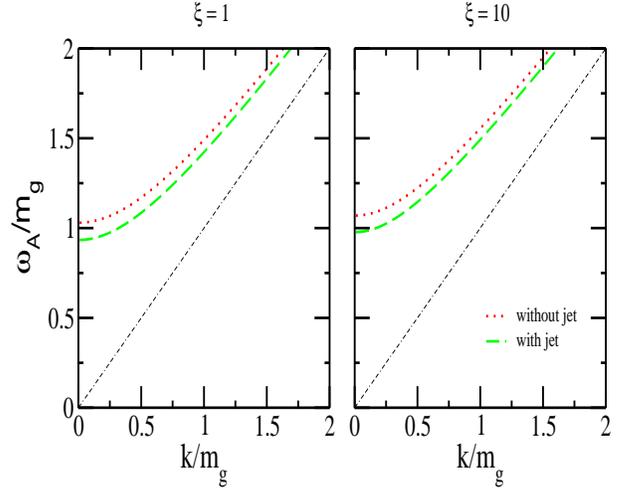}
\end{center}
\caption{(Color online) The dispersion relation for the stable A-mode for an anisotropic plasma 
with jet for different values of the anisotropy parameter $\xi=\{1, 10\}$, $\eta=0.2$, $\theta_{jet}=\pi/2$,
$v_{jet}=0.7$ and $\theta_n=0$}  
\label{fig1}
\end{figure}
\begin{figure}[t]
\begin{center}
\epsfig{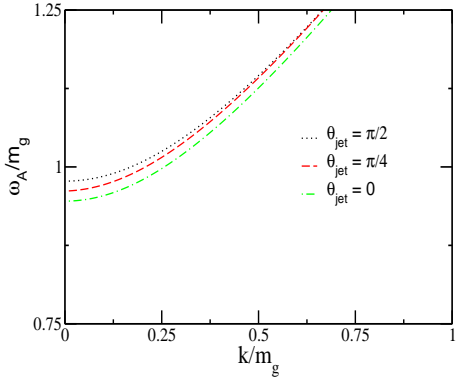}
\end{center}
\caption{(Color online) Same as Fig.\protect\ref{fig1}
for  $\xi=10$, $v_{jet}=0.7$, $\theta_{n}=0$, $\eta=0.2$ and $\theta_{jet}=\{0,~\pi /4,~\pi /2\}$.}  
\label{fig2}
\end{figure}
\begin{figure}[t]
\begin{center}
\epsfig{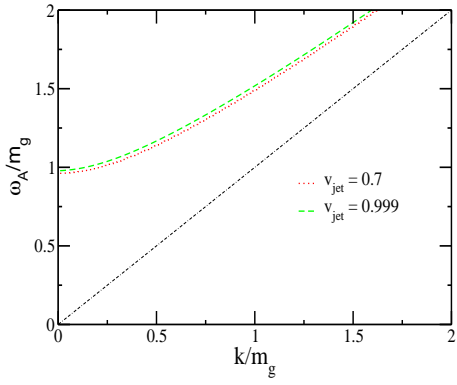}
\end{center}
\caption{(Color online) Same as Fig.\protect\ref{fig1} for
$v_{jet}=\{0.7,~0.999\}$,  $\xi=10$, $\theta_{n}=0$ and $\theta_{jet}=\pi/4$.}  
\label{fig3}
\end{figure}
\begin{figure}[t]
\begin{center}
\epsfig{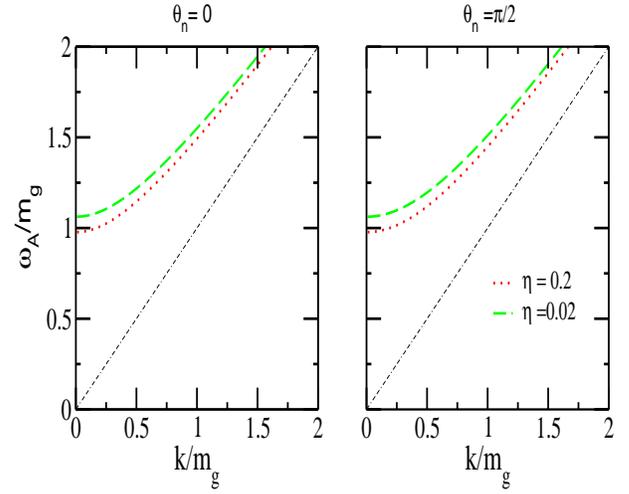}
\end{center}
\caption{(Color online) The dispersion relation for the stable A-mode for the system composed of anisotropic plasma 
and  jet for $\xi=10,~v_{jet}=0.7,~ \theta_{jet}=\pi/2$ and $\eta=\{0.02,~0.2\}$. 
The left(right) panel corresponds to  $\theta_{n}=0(\pi/2)$.}
\label{fig4}
\end{figure}
\begin{figure}[t]
\begin{center}
\epsfig{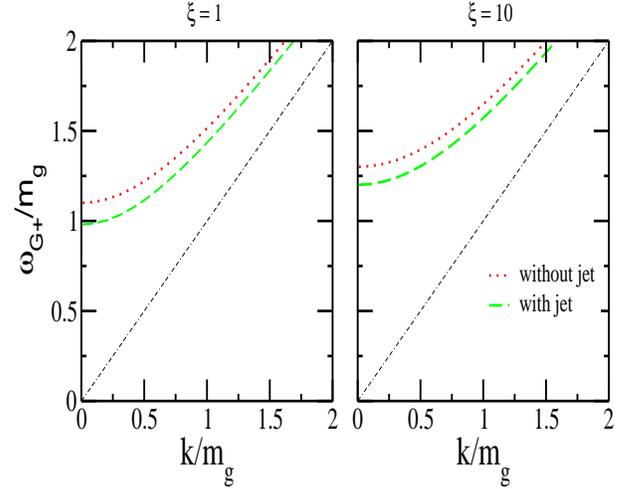}
\end{center}
\caption{(Color online) The dispersion relation for the stable $G_+$-mode for an anisotropic plasma 
with jet  for $\theta_n=\pi/2$. The left(right) panel corresponds to anisotropy parameter $\xi=1(10)$.}  
\label{fig5}
\end{figure}
\begin{figure}[t]
\begin{center}
\epsfig{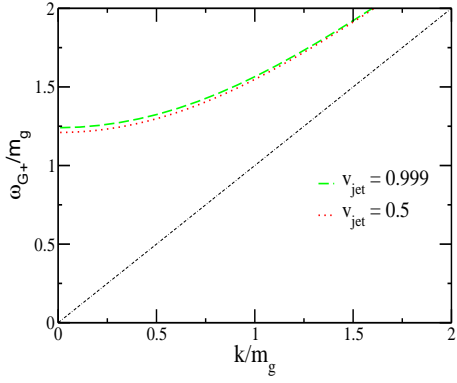}
\end{center}
\caption{(Color online) Same as Fig.\protect\ref{fig5}
for $v_{jet}=\{0.5,~0.999\}$, $\xi=10,$ $\theta_n=\pi/3$, $\theta_{jet}=\pi/2$ and $\eta=0.2$.}  
\label{fig7}
\end{figure}
\begin{figure}[t]
\begin{center}
\epsfig{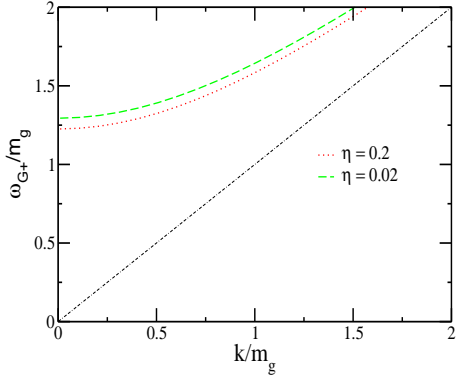}
\end{center}
\caption{(Color online) Same as Fig.\protect\ref{fig5}
for $\eta=\{0.02,~0.2\}$, $\xi=10,$ $v_{jet}=0.7$, $\theta_{n}=\pi/2$ and $\theta_{jet}=\pi/2$.}  
\label{fig8}
\end{figure}
\begin{figure}[t]
\begin{center}
\epsfig{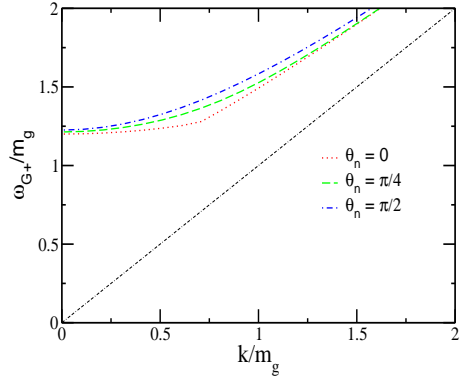}
\end{center}
\caption{(Color online)  Same as Fig.\protect\ref{fig5}
for $\theta_{n}=\{0, \pi /4,  \pi /2\}$, $\xi=10,$ $v_{jet}=0.7$, $\theta_{jet}=\pi/2$ and $\eta=0.2$.}  
\label{fig9}
\end{figure}

These solutions are found numerically by varying the various parameters introduced earlier. The results for  the stable $A$-mode are 
shown in Fig.\ref{fig1} for two values of the anisotropic parameter $\xi=\{1,~10\}$, $\eta=0.2$, $\theta_{jet}=\pi/2$, 
$v_{jet}=0.7$ and $\theta_{n}=0$ with and without the jet. It is seen that the collective modes with the jet differs reasonably
from that without the jet. 
Next we consider the dependence of the dispersion relation on 
the angle of propagation of the jet with the wave vector. This is 
displayed in Fig.\ref{fig2} for $\xi=10$, $v_{jet}=0.7$, $\theta_{n}=0$ and $\eta=0.2$. It is found that the dispersion
relation is sensitive to $\theta_{jet}$. In order to see the dependence of the collective modes on the jet velocity
$v_{jet}$, we plot the dispersion relation for A-mode for fixed $\theta_{n}$ and $\theta_{jet}$ in Fig.\ref{fig3} for two values 
of $v_{jet}$. The dispersion relation is modified again, but it is
not that sensitive with $v_{jet}$.
The dispersion relation for the stable A-modes in the composite system 
is shown in the Fig.\ref{fig4} for $\theta_n=0,~{\rm and}~\pi/2$. We have  fixed the parameters as explained in the 
figure. It is observed that for a fixed value of $\theta_n$ the modes have significant dependence on the value of $\eta$.
When the wave vector is parallel to the direction of the anisotropy, i.e., 
${\bf k}||{\bf \hat{n}}$ then $\gamma$ and $\tilde{n}^2$ vanish identically. 
Therefore  the dispersion relations for $G_+$-modes are similar to $A$-modes for $\theta_n=0$.  
The dependence of the $G_+$-modes on the anisotropic parameter $\xi$ is displayed in Fig.\ref{fig5}. For fixed $\theta_n$ we
find significant difference in the collective modes with and without jet. 
However we do not find any significant dependence of the $G_+$-modes on the jet 
velocity(see in Fig.\ref{fig7}).
From  Fig.\ref{fig8}, it is clearly seen that the dispersion relation depends on the parameter $\eta$ and the modes are 
strongly modified by this parameter.
The collective modes also depend on the angle of propagation with respect to the anisotropy vector, $\theta_n$ as
depicted in Fig.\ref{fig9}.  
The numerical results for the stable $G_-$-mode are shown in Fig.\ref{fig10} for  $\xi=\{1,~10\}$, $\eta=0.2$, 
$\theta_{jet}=\pi/2$, $v_{jet}=0.7$ and $\theta_{n}=0$ with and without jet. We find  significant difference in the 
collective modes with and without jet and with jet $G_-$-mode quickly bends towards the $\omega=k$ line. 
The dispersion relation for $G_-$ mode for $\theta_{jet}=\{0,\pi/2\}$ is displayed in Fig.\ref{fig11}. Unlike the 
$A$ and $G_+$-modes, it is seen that the dispersion relation for $G_-$-mode is more sensitive to $\theta_{jet}$.
Like the $A$ and $G_+$-modes, the collective mode for $G_-$-mode is also 
marginally sensitive to the jet velocity. In Fig.\ref{fig12}, it is observed that the 
dispersion relation for $G_-$-mode strongly depends on the direction of the anisotropy axis. 
\newline

\begin{figure}[t]
\begin{center}
\epsfig{file=n1a-d.eps,width=8cm,height=6.5cm,angle=0}
\end{center}
\caption{(Color online) The dispersion relation for the stable $G_-$-mode for an anisotropy plasma and 
a jet for $\theta_n=\pi/2$. The left(right) panel corresponds to anisotropy parameter $\xi=1(10)$.}  
\label{fig10}
\end{figure}

\begin{figure}[t]
\begin{center}
\epsfig{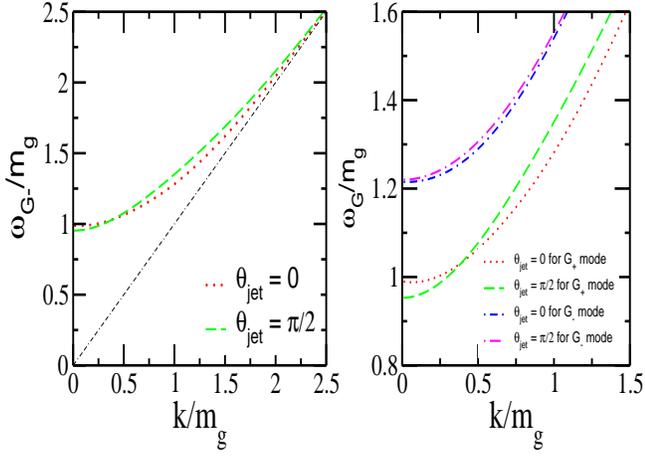}
\end{center}
\caption{(Color online) The left(right) panel corresponds to the 
dispersion relation for the stable $G_-$(G)-mode for the composite system 
for  $\theta_{jet}=\{0,~\pi /2\}$, $\xi=10$, $v_{jet}=0.7$, $\theta_{n}=\pi/3$ and $\eta=0.2$.}  
\label{fig11}
\end{figure}

\begin{figure}[t]
\begin{center}
\epsfig{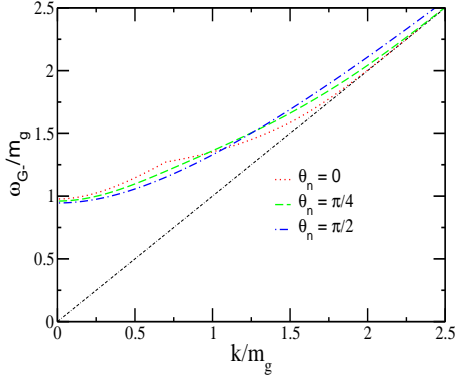}
\end{center}
\caption{(Color online) Same as Fig.\protect\ref{fig10}
for $\xi=10$, $v_{jet}=0.7$, $\theta_{jet}=\pi/2$, $\eta=0.2$ and different $\theta_{n}=\{0, \pi /4,  \pi /2\}$.}  
\label{fig12}
\end{figure}

\subsection{Unstable modes}
In the static limit, for small $\theta_n$ both the scale $m^2_{\alpha}$ and $m^2_-$ are negative and so also 
$\alpha^{\prime}$. It indicates that the  whole system is unstable with 
respect to magnetic  instability~\cite{bjp,jhep08}. This can be identified
as the so called filamentation or Weibel instability~\cite{prl2}. The instability is driven by the energy transferred from the 
particles  to the field. 
The growth rate  of  the filamentation instabilities is  largest when the  wave vector is along the beam line, 
i.e., ${\bf k}||{\bf \hat{n}}(\theta_n=0)$~\cite{prd68,prd70,prd73} in which case  
$\gamma$ and $\tilde{n}^2$ vanish identically.
Therefor the dispersion relations for the unstable modes 
are determined by the solutions of the following equations:
\bea
\omega^2-{\bf k}^2-\alpha(\omega)-\alpha^{\prime}(\omega)&=&0\\
\omega^2-\beta(\omega)-\beta^{\prime}(\omega)&=&0
\eea  

\begin{figure}
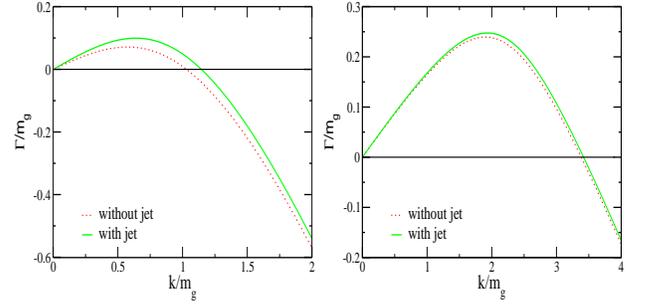

\begin{center}
\epsfig{file=u1a-b.eps,width=4cm,height=4cm,angle=0}~\epsfig{file=u2.eps,width=4cm,height=4cm,angle=0}
\end{center}
\caption{(Color online) The growth rate $\Gamma$ of the unstable $\alpha-\alpha^{\prime}$ mode, for two different anisotropy parameter 
$\xi=\{1, 10\}$ and $\theta_{jet}=0$.}  
\label{figu1}
\end{figure}

In the numerical simulation we find that unstable mode exits only for $\alpha-\alpha^{\prime}$ mode
in the special case$({\bf k}||{\bf {\hat{n}}})$ considered here. 
To solve the dispersion relation for the unstable $\alpha-\alpha^{\prime}$ 
mode  we first consider $\theta_{jet}=0$ and it is found that $\omega$ is
purely imaginary i.e. $\omega=i\Gamma$ with $\Gamma$ real valued. 
Results for different values of the anisotropy parameter 
$\xi=\{1,~10\}$ are shown in Fig.\ref{figu1}. 
It is observed that instability first increases in comparison to 
the no jet case and then it becomes damped.
To find the maximum value of the momentum $k_{max}$ at which the unstable mode spectrum terminates, we take the limit 
$\Gamma\rightarrow0$, to obtain($\theta_{jet}=0$), 
\bea
k^2_{max} = \omega^2_{jet}+m_D^2\frac{\sqrt{\xi}+(\xi-1)\arctan(\sqrt{\xi})}{4\sqrt{\frac{\xi}{\xi+1}}}.
\eea

\begin{figure}
\begin{center}
\epsfig{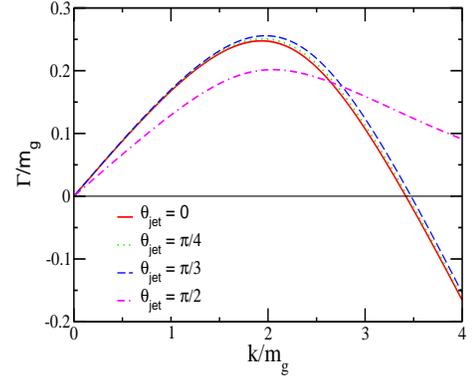}
\end{center}
\caption{(Color online) Same as Fig.\protect\ref{figu1} with 
$\theta_{jet}=\{0, \pi/4, \pi/3, \pi/2\}$, $\xi=10$ and $\eta=0.2$.}
\label{figu2}
\end{figure}

We shall now consider $\theta_{jet}$ dependence
of the unstable modes. For nonzero $\theta_{jet}$
we find that collective modes are unstable propagating modes i.e.  
$\omega=a+i\Gamma$. In Fig.\ref{figu2} we plot  the imaginary part of the 
dispersion law of the unstable $\alpha-\alpha^{\prime}$-mode for four different 
$\theta_{jet}$. It is clearly seen that  $\Gamma$ first increases with 
$\theta_{jet}$ and then turns into damping except $\theta_{jet}=\pi/2$
in which case the unstable mode always grows. 
It is interesting to see that when $\theta_{jet}=\pi/2$, the maximum value 
of $\Gamma$ decreases and it  never becomes damped.  

\begin{figure}
\begin{center}
\epsfig{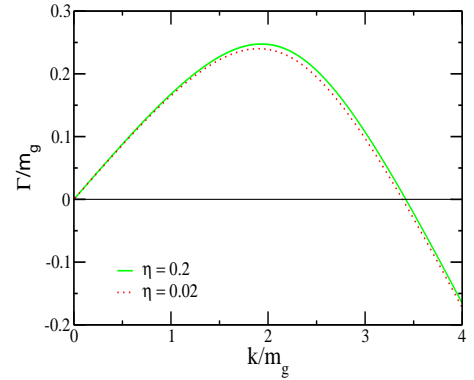}
\end{center}
\caption{(Color online) Same as Fig.\protect\ref{figu1} with 
$\eta=\{0.02, 0.2\}$, $\xi=10$ and $\theta_{jet}=0$.}
\label{figu3}
\end{figure}

The growth rate dependence on the parameter $\eta=\{0.02,~0.2\}$ corresponding to the 
value of the $\xi=10$ and $\theta_{jet}=0$ is displayed in Fig.\ref{figu3}. 
It is seen that instability grows with $\eta$. 


\section{Summary}
We have  calculated the dispersion relation of modes (both stable and unstable) due to the passage of relativistic jets
in an anisotropic plasma. It is seen that the dispersion relations of stable modes in an AQGP are modified due to the 
introduction of relativistic jet compared to the case when there is no jet. Dependences of possible modes on the direction of
propagation of the jet and anisotropy direction have been demonstrated. We have also shown that the growth rate $\Gamma$ 
increases with the passage of jet and this growth rate strongly depends on the anisotropy parameter $\xi$. This means that 
the introduction of jet in an AQGP leads to faster isotropization for the special case $(\hat{k}||\hat{n})$ 
considered here. It is also interesting to note 
that the growth rate for $\theta_{jet}=\pi/2$ never becomes damped.
We also find no unstable for $\beta-\beta^{\prime}$-mode  in the special case.
However, it remains to be seen how the unstable modes behave when
a more general case(arbitrary $\theta_n$) is considered.


\begin{thebibliography}{50}


\medskip


\bibitem{prd74} C.~Manuel and S.~Mrowczynski, Phys.~Rev.~D {\bf 74}, 105003 (2006).
\bibitem{prd76} M.~Mannarelli and C.~Manuel, Phys.~Rev.~D {\bf 76}, 094007 (2007).  
\bibitem{prd77} M.~Mannarelli and C.~Manuel, Phys.~Rev.~D {\bf 77}, 054018 (2008).
\bibitem{hep-ph971} R.~D.~Pisarski, hep-ph/9710370
\bibitem{prc49} S.~Mrowczynski, Phys.~Rev.~C {\bf 49}, 2191 (1994).
\bibitem{prl94} P.~Arnold, J.~Lenaghan, G.~D.~Moore and L.~G.~Yaffe Phys. Rev. Lett. {\bf 94}, 072302 (2005)
\bibitem{prl94A} A.~Rebhan, P.~Romatschke and M.~Strickland Phys. Rev. Lett. {\bf 94} 102303 (2005)
\bibitem{appb} S.~Mrowczynski, Acta Phys.Polon. {\bf B 37} 427 (2006).


\bibitem{prd68} P.~Romatschke and M.~Strickland Phys. Rev. D {\bf 68}, 036004 (2003).
\bibitem{prd70} P.~Romatschke and M.~Strickland Phys. Rev. D {\bf 70}, 116006 (2004). 
\bibitem{pr183} H.~T.~Elze and U.~Heinz Phys. Rep. {\bf 183}, 81 (1989).
\bibitem{prl2} E.~Weibel Phys. Rev. Lett. {\bf 2}, 83 (1953).
\bibitem{prd59} P.~Arnold, D.~T.~Son and L.~G.~Yaffe Phys.Rev.D {\bf 59} 105020 (1999)
\bibitem{prd73} B.~Schenke, M.~Strickland, C.~Greiner and M.~H.~Thoma Phys. Rev. D {\bf 73}, 125004 (2006).
\bibitem{plb662} A.~Dumitru, Y.~Guo, and M.~Strickland, Phys. Lett. {\bf B662}, 37 (2008).
\bibitem{bjp} M.~Strickland Brazilian Journal of Physics, vol. 37, no. 2C (2007).
\bibitem{jhep08} P.~Arnold, J.~Lenaghan and G.~D.~Moor  JHEP {\bf 08}, 002 (2003); 
P.~Arnold, G.~D.~Moor and L.~G.~Yaffe, JHEP {\bf 01}, 030 (2003). 



\end{thebibliography}
\end{document}